\documentclass[conference]{IEEEconf}
\usepackage{balance}
\usepackage{color}
\usepackage{fancybox}
\usepackage{graphicx}
\usepackage{cellspace}
\usepackage{comment}
\usepackage{diagbox}
\usepackage{url}
\usepackage{xspace}
\usepackage{amssymb}
\usepackage{multirow}
\usepackage[caption=false]{subfig}
\usepackage{epsfig}
\usepackage{cellspace}
\usepackage[english]{babel}
\usepackage[utf8x]{inputenc}
\usepackage[T1]{fontenc}
\usepackage{amsmath}
\usepackage{tikz}
\usepackage{xcolor}
\graphicspath{{images/}}
\usepackage{adjustbox} 

\usepackage{titlesec}
\usepackage{titletoc}

\graphicspath{{images/}}

{\begin{list}{}%
        {\setlength{\leftmargin}{#1}}%
       \item[]%
}
{\end{list}}

\newcommand{\ie}{{\textit{i.e.,}}}
\newcommand{\al}{\textit{et al.\xspace}}

\begin{document}

\title{SPIRT: A Fault-Tolerant and Reliable Peer-to-Peer Serverless ML Training Architecture}






\author{
\IEEEauthorblockN{
Amine BARRAK\IEEEauthorrefmark{1},
Mayssa JAZIRI\IEEEauthorrefmark{1},
Ranim TRABELSI\IEEEauthorrefmark{1},
Fehmi JAAFAR\IEEEauthorrefmark{1},
Fabio PETRILLO\IEEEauthorrefmark{2}
}
\IEEEauthorblockA{\IEEEauthorrefmark{1}\textit{Department of Computer Science and Mathematics}, \textit{University of Quebec at Chicoutimi, UQAC}, Saguenay, Canada \\
Email: \{mabarrak, mjaziri, ranim.trabelsi1, fehmi.jaafar\}@uqac.ca}
\IEEEauthorblockA{\IEEEauthorrefmark{2}\textit{Département de génie logiciel}, \textit{École de Technologie Supérieure, ÉTS}, Montreal, QC \\
Email: fabio.petrillo@etsmtl.ca}
}

\maketitle

\begin{abstract}

The advent of serverless computing has ushered in notable advancements in distributed machine learning, particularly within parameter server-based architectures. Yet, the integration of serverless features within peer-to-peer (P2P) distributed networks remains largely uncharted. In this paper, we introduce SPIRT, a fault-tolerant, reliable, and secure serverless P2P ML training architecture. designed to bridge this existing gap.

Capitalizing on the inherent robustness and reliability innate to P2P systems, SPIRT employs RedisAI for in-database operations, leading to an 82\% reduction in the time required for model updates and gradient averaging across a variety of models and batch sizes. This architecture showcases resilience against peer failures and adeptly manages the integration of new peers, thereby highlighting its fault-tolerant characteristics and scalability.
Furthermore, SPIRT ensures secure communication between peers, enhancing the reliability of distributed machine learning tasks. Even in the face of Byzantine attacks, the system's robust aggregation algorithms maintain high levels of accuracy. These findings illuminate the promising potential of serverless architectures in P2P distributed machine learning, offering a significant stride towards the development of more efficient, scalable, and resilient applications.

\end{abstract}
\vspace{0.1cm}

\section{Introduction}
\label{sec:introduction}

Distributed machine learning (ML) has become increasingly important in tackling the rising complexity and large amounts of data associated with ML models, which traditional single-machine learning frameworks struggle to manage. This approach harnesses multiple computational nodes for simultaneous processing, greatly reducing the time necessary to train larger, more intricate models\cite{DistributedNN}. 

Multiple distributed ML architectures have been introduced over the years, many of which are fundamentally based on the structures of the Parameter Server (PS) and Peer-to-Peer (P2P) architectures \cite{verbraeken2020survey}. Each of these represents a unique methodology for orchestrating the management and distribution of tasks and data across nodes in a distributed system, with their own set of benefits and challenges \cite{alqahtani2019performance, sun2022decentralized}.

In the parameter server architecture, for instance, the worker nodes perform computations on their respective data partitions and communicate with the parameter server (PS) to update the global model \cite{li2013parameter}. In contrast, peer-to-peer (P2P) architectures distribute the model parameters and computation across all nodes in the network, eliminating the need for a central coordinator \cite{vsajina2023peer}. 

In the pursuit of optimizing distributed ML architectures, serverless computing has emerged as a revolutionary development. Today, the growing trend of serverless computing, including platforms such as Amazon Lambda\cite{Serverle24:online}, Google Cloud Functions\cite{CloudFun20:online}, and Azure Functions\cite{CloudCom14:online}, offers flexibility and cost-effective solutions for distributed ML systems\cite{barrak, P52, P15}. A common trait among existing serverless architectures—whether based on Parameter Server or P2P setups—is a heavy reliance on databases as a communication channel. This necessity arises from the stateless nature of serverless architectures and limitations in directly transmitting large data sizes\cite{P53, P46}. As a result, communication latency can arise, especially during the iterative process of model update and gradient aggregation\cite{P46, chahal2022pay}. In fact, in these architectures, the model parameters and computed gradients must be fetched from the database, processed, and re-uploaded in each iteration. This recurring cycle imposes a significant burden, leading to additional computational and communication costs that can significantly impact overall efficiency.

Notably, while the usage of serverless computing within parameter server architecture has been extensively studied and proven to yield benefits like cost reduction \cite{sarroca2022mlless}, scalability \cite{P54,P39}, and performance efficiency \cite{sampe2018serverless}. Surprisingly, however, the potential of serverless computing within peer-to-peer (P2P) architectures remains less explored, despite the immense advantages P2P offers in terms of data privacy, load balancing, and fault tolerance, as a single node's failure does not disrupt the system \cite{anthony2015systems}.

In relation to this, our preceding work \cite{ic2e2023_barrak} explored the integration of serverless computing within VM-based P2P architecture, aiming to offload excessive gradient computation when faced with resource constraints. Despite the benefits of this strategy, it represents a piecemeal approach that does not fully harness the potential of a completely serverless system.

Adding to these challenges, security is a significant concern in distributed Machine Learning (ML) environments, particularly in Peer-to-Peer (P2P) based architectures \cite{wang2023sparsfa}. One issue is the absence of stringent mechanisms to authenticate new peers joining the network \cite{xu2022spdl, shayan2018biscotti}. Unauthenticated or malicious peers could join the network, potentially causing significant disruption to the training process.

Beyond the threat of unauthenticated new peers, even existing, trusted peers can become compromised, introducing and contributing malicious gradients intended to sway the model training in harmful directions, severely compromising the model's integrity and performance \cite{guerraoui2021garfield, xu2022spdl, fang2019bridge}. This risk emphasizes the importance of robust aggregation methods \cite{blanchard2017machine, xie2018generalized, xie2019zeno}, which can help mitigate the effect of such outliers on the overall model's performance.

Addressing the outlined challenges, we introduce the Serverless Peer Integrated for Robust Training (SPIRT) Architecture. In this setup, each peer is identified by its associated database, e.g., Redis, and executes shard-based gradient computation, aggregation, and model convergence checks. With orchestrated workflow coordination, e.g., AWS Step Functions, SPIRT effectively overcomes the stateless limitations inherent in serverless computing. Consequently, it takes full advantage of the serverless architecture within a peer-to-peer setup for machine learning training.

In this endeavor, we propose a reliable peer to peer, serverless ML training architecture with the following contributions:

\begin{enumerate}
    \item Automated ML training Workflow: We propose an automated ML training serverless workflow within a P2P architecture, maximizing efficiency and scalability while reducing operational complexities.

    \item Serverless-adapted Redis Enhancement: We modify RedisAI for serverless ML training systems that rely on databases, introducing in-database model updates to eliminate the traditional fetch-process-reupload cycle. This enhancement streamlines communication and reduces overhead, facilitating more efficient serverless ML operations.
    
    \item Enhanced Fault Tolerance: We integrate secure, scalable peer-to-peer communication, new participant integration, and robust aggregation mechanisms to establish a fault-tolerant environment. This approach ensures reliable distributed ML training, even in the presence of potential disruptions or adversarial actors.
\end{enumerate}

Once the paper is accepted, a replication package will be made available for further research.


\section{Related work}
\label{sec:relatedwork}
This section reviews literature on distributed machine learning and serverless computing, focusing on decentralized ML on P2P paradigm and fault tolerance mechanisms.

\subsection{Distributed Machine Learning Architectures}
Distributed machine learning has gained prominence due to the intense computational requirements and extensive data involved in training sophisticated ML models \cite{haussmann2018accelerating, verbraeken2020survey, zaharia2010spark}. The choice of topology, either Centralized, Tree-like or P2P, is a critical factor influencing performance, scalability, reliability, and security \cite{baran1964distributed, verbraeken2020survey, agarwal2014reliable, wei2015managed}. P2P architectures, for instance, offer scalability and reliability by eliminating the "Single Point of Failure". Recent studies have begun to explore serverless computing for machine learning, a promising yet nascent field \cite{barrak, P15, P34, P54, P53}. While many works \cite{P51, P53, P49, P46, P39, P38} are based on the PS paradigm, others have started to present decentralized learning solutions using serverless computing \cite{ic2e2023_barrak}. 

Unlike previous studies, we propose a fully automated ML training serverless workflow within a P2P architecture.
 
\subsection{Communication and Model Updating in Serverless based Distributed ML}
Serverless computing coupled with database systems has seen increased application in the field of distributed learning. Works such as LambdaML \cite{P46}, SMLT \cite{P54}, and MLLess \cite{sarroca2022mlless} utilize a database as a communication channel, collecting intermediate local statistics from all workers through a designated communication pattern. Other works propose in-database distributed machine learning solutions to reduce communication overhead \cite{10.14778/3352063.3352083}. Additionally, libraries such as RedisAI \cite{RedisAIA54:online} enable seamless inference on trained models within Redis, and other solutions like OML4R \cite{OracleRe26:online} and MADlib \cite{hellerstein2012madlib} offer SQL-based functions for prediction and inference within database environments. 

Differentiating from these works, our architecture utilizes Redis, a high-performance in-memory data structure store, serving as both a communication channel and an automatic model update tool through RedisAI.

\subsection{Security and Robustness in Distributed ML}
Fault tolerance in distributed ML is facilitated by techniques like Availability Zones, retries, architectural decisions, and checkpointing mechanisms \cite{Crossreg87:online,Configur50:online, qiao2019fault, P53, P46}. The heartbeat technique is also used for failure detection \cite{bouizem2020active}. Several studies also propose securing communication within distributed systems to prevent data leakage \cite{P39, wink2021approach, xu2022spdl, bourreau2022securing}. In addition, robust aggregation techniques such as KRUM \cite{blanchard2017machine}, MULTI-KRUM \cite{blanchard2017machine}, GeoMed \cite{xie2018generalized}, MarMed \cite{xie2018generalized}, and ZENO \cite{xie2019zeno} are used to address Byzantine faults \cite{lamport2019byzantine}. These robust aggregations rules have been adapted to P2P architectures in works like the BRIDGE framework \cite{fang2019bridge} and a blockchain-based solution by Xu \al~\cite{xu2022spdl}. 

We expand on this body of work by implementing a secure, scalable peer-to-peer communication, new participant integration into the ML training network, and incorporating robust aggregation mechanisms for reliable distributed ML.

\begin{figure*}[ht]
 \centering
 \includegraphics[scale=0.48]{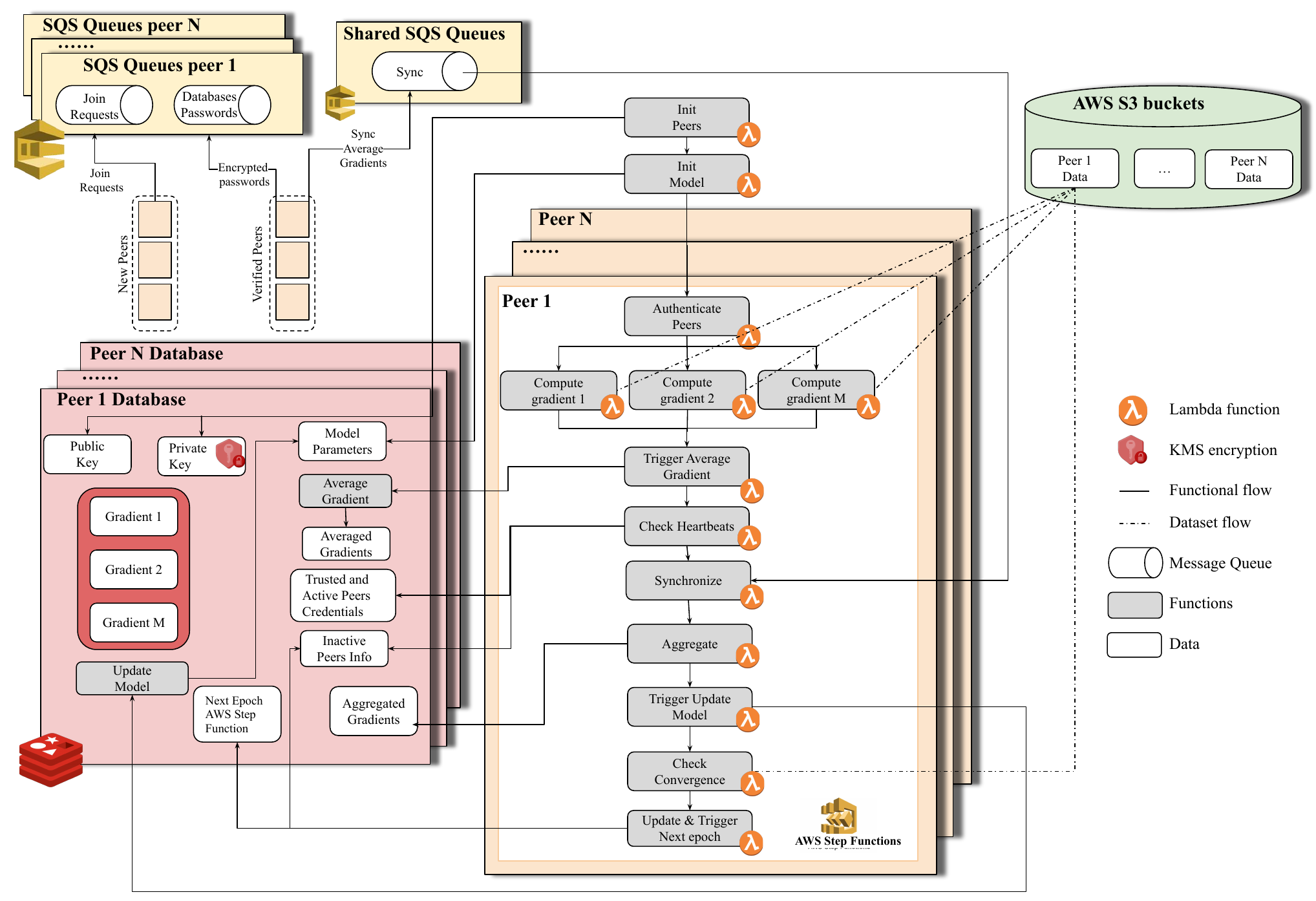}
 \caption{Overview of the proposed Peer To Peer training based on Serverless computing}
 \label{fig:arch}
 \vspace*{-10pt}
\end{figure*}

\section{Design Architecture of Logical Peer to Peer training ML}
\label{sec:study-design}
In this section, we delve into the design architecture of a serverless, peer-to-peer machine learning training system. The discussion encompasses a broad overview of the proposed architecture, an in-depth examination of the core components, and an exploration of the operational dynamics driving the system's overall functionality and performance.

\subsection{Comprehensive Overview of the Proposed Architecture}

Our proposed serverless P2P distributed training architecture, as illustrated in Figure \ref{fig:arch}, kicks off with system initialization, followed by the authentication of new peers, if any. In this configuration, every peer in the network is uniquely identified by the IP address and port tied to its corresponding stateful component - a dedicated Redis database. A heartbeat monitoring system ensures the constant availability of all peers.

Within the system, the assigned dataset of each peer is divided into smaller shards (batches). The peer computes gradient for each shard, averages them, and stores the result in its Redis database. These averaged gradients are then collectively aggregated among all peers, filtering out any outliers in the process by applying robust aggregation. The resultant set of trusted, aggregated gradients is used to update the model parameters. The system performs periodic checks for model convergence \ie every ten epochs, assuring optimal progress during the learning phase. Data integrity and confidentiality are ensured through secure peer communication.

The entire process is coordinated using AWS Step Functions, which seamlessly orchestrates the flow of each epoch within the training process of each peer.

This architecture is deployed on Amazon AWS for its unique benefits like the 15-minute timeout and 10GB RAM from AWS Lambda \cite{Lambdaqu81:online}. Notably, comparable services on platforms like Google Cloud, Azure, and IBM Cloud enable possible architecture replication.

\subsection{Deep Dive into Core Architectural Components}
\label{deep}
Following the initial overview, we will now go over each key facet of our proposed Peer-to-Peer (P2P) serverless architecture.

\subsubsection{\textbf{Training Dataset Management and Partitioning}}

Individual peers have the ability to pull data from multiple distributed storage systems, including Amazon S3. The specific data each peer is responsible for is determined by its unique rank. This data is then divided into smaller units, or shards, to enable batch processing.

\subsubsection{\textbf{Leveraging Serverless Computing Across Peer Training Tasks}}

The cornerstone of our architecture is serverless computing, embodied by Amazon Lambda functions. Incorporated throughout the peer training workflow—from peer authentication to model updates—it offers benefits such as isolation for uninterrupted operations, capacity for managing compute-intensive tasks like gradient computations, and scalability for workload fluctuations.

\subsubsection{\textbf{Training Workflow Orchestration}}
Due to the serverless structure of our system, where functions operate independently, their orchestration is crucial. To manage this, we use AWS Step Functions, a powerful serverless workflow service. It coordinates the entire machine learning training process within each peer during each epoch, including tasks like peers authentication, gradient computation and averaging, model updates, and convergence assessment. Notably, our architecture integrate continuous invocation of step functions for each epoch, which effectively mitigates AWS Lambda's cold start delays, thereby enhancing overall performance and minimizing latency during ML training.

\subsubsection{\textbf{State Management and Processing in Database}}

In our architecture, we use Redis, an open-source in-memory data store, for quick access to machine learning artifacts such as model parameters and gradients - a key requirement for any stateless distributed ML system. 

Beyond a simple key-value store, we utilize the RedisAI module which supports various deep learning backends, enabling in-database ML operations, and minimizing data transfer latency. RedisAI is especially efficient at serving models at scale and in real-time. Unique to our architecture is the extension of RedisAI's capabilities to directly modify model parameters within the database, eliminating the traditional process of external processing, our routine performs these operations inside the database itself.

\subsubsection{\textbf{Synchronization between Peers}}

Within our proposed architecture, achieving synchronization amongst peers is paramount for ensuring the correctness of the distributed training process. To manage this aspect of distributed computation, we employ the AWS Simple Queue Service (SQS).

Once a peer completes gradient computation for its data shards and averages local gradients, it sends a notification message to a designated synchronization queue, the ``Sync Queue'', signifying the task completion. If a peer doesn't respond or acknowledge within a designated timeout period, others proceed without waiting indefinitely. The unresponsive peer is identified as a failed node in the next epoch by our heartbeat monitoring system.

We note that the messages inside the ''sync queue`` will be deleted by any peer in initialisation phase.

\subsubsection{\textbf{Secure Communication: Safeguarding Data Integrity and Confidentiality}}

Our architecture employs stringent secure communication protocols to ensure data integrity and confidentiality during inter-peer interactions, accomplished through the RSA algorithm for asymmetric encryption, with unique public and private keys for each peer. Beyond encryption, unique digital signatures derived from private keys authenticate sender identity and verify data integrity.

Each peer's private key is safeguarded by encryption using a unique key from AWS Key Management Service (KMS), with access strictly limited to few authorized services (Lambda functions), enhancing security against unauthorized access.

\subsection{Operational Dynamics of the Proposed Architecture}
Within this subsection, we take a closer look on the operational dynamics that power our proposed architecture. Moving beyond the standalone examination of the key components, to their interactions, collaborative processes, and the mechanisms that drive the overall system performance and functionality.

\subsubsection{\textbf{Peers Initialization and Authentication}}

Our architecture relies on Amazon SQS for two main operations: \textit{peers' initialization} and \textit{new peers' integration}. Each peer has two distinct SQS queues - the first for join requests and the second for receiving encrypted passwords of other peers' databases. Every peer also maintains an AWS Key Management Service (KMS) encryption key, securing its private key within its database, and ensuring exclusive key access.

\textbf{Peers Initialisation}

\begin{figure}[ht]
 \centering
 \includegraphics[scale=0.45]{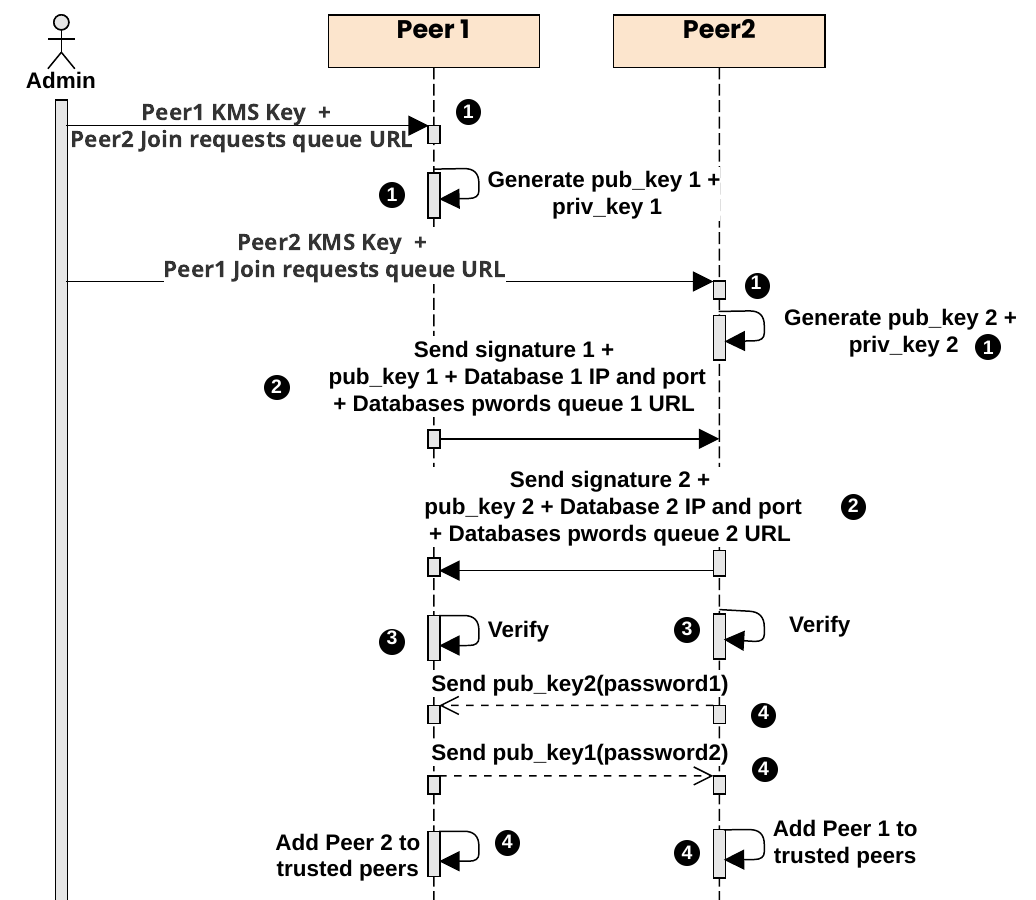}
 \caption{Sequence Diagram Illustrating the Initialization Process of Peers}
 \label{fig:peer_init}
\end{figure}

This phase involves the setup of individual peers. The process below outlines an example of initializing a two-peer system, and it's also illustrated in Figure \ref{fig:peer_init}:

\begin{enumerate}
\item At the onset, the admin initiates the peers, by providing each their own KMS encryption key, URLs of their neighboring peers' ``join requests'' SQS queues, and a unique rank. Each peer then generates private and public keys. The public key is stored in its plain form, while the private key is encrypted using the KMS encryption key before storage in the respective database.

\item Each peer then generates a digital signature and broadcasts this, along with its public key, database IP address and port, and the URL of its ``databases passwords'' queue inside other peer's ``join requests'' queue.

\item Verification ensues as each peer validates the other's signature
\item Upon successful verification, peers mutually exchange their encrypted database passwords. Furthermore, they save each other details, including the rank, into their respective databases. 

\end{enumerate}

\begin{figure}[ht]
 \centering
 \includegraphics[scale=0.43]{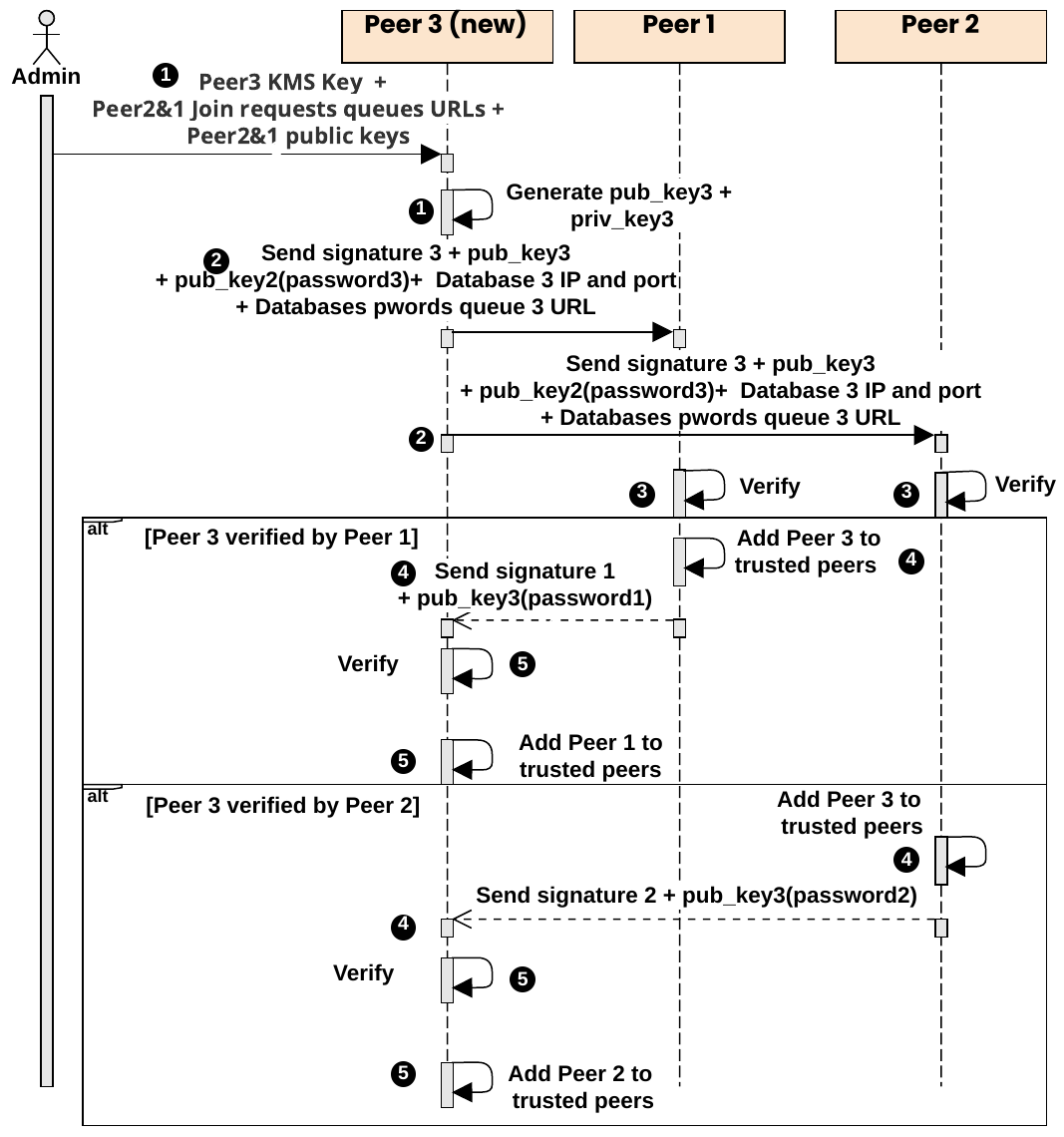}
 \caption{Sequence Diagram of New Peer Integration into Training Network}
 \label{fig:new_peer}
  \vspace*{-10pt}
\end{figure}

\textbf{Novel peer integration}
In this phase, we illustrate the process of integrating a new peer into an existing training network of two peers, as shown in Figure \ref{fig:new_peer}. The steps are as follows:

\begin{enumerate}
\item The process is initiated when the admin provisions the new peer (Peer 3) with the URLs of the ``join requests'' SQS queues of existing peers and their corresponding public keys, and their ranks. Peer 3, in response, generates a pair of public and private keys. These keys are stored in its database, with the private key encrypted using the designated KMS encryption key.

\item The new peer (Peer 3) generates a digital signature, broadcasting it and its public key, database port and IP address, URL of its ``databases passwords'' queue, and rank. It sends its password, encrypted with the recipient's public key, via their ``join request'' queues.

\item Once the broadcasting phase is completed, Peer 3 waits for validation from the existing peers. These peers undertake the task of validating Peer 3's authenticity. They accomplish this by rigorously comparing the signature provided by Peer 3 against the information contained in its public key, ensuring a match.

\item  After Peer 3's successful validation, Peers 1 and 2 send back their individual signatures, and their databases passwords encrypted with Peer 3's public key, into Peer 3's ``databases passwords'' queue. Furthermore, they incorporate Peer 3's details into their respective databases.

\item Finally, Peer 3 validates the sender peers based on their signatures and public keys. Upon successful validation, Peer 3 records the details of Peers 1 and 2 into its own database.
\end{enumerate}

\subsubsection{\textbf{Model Initialisation}}
The model initialization phase involves the establishment of a unified model to serve all initialized peers. This model can be initialized with random parameters or pre-trained ones, depending on the requirements. The chosen model, which could be a specific ML or deep learning model, is then stored in each peer's Redis database using RedisAI. This ensures a consistent starting point for distributed learning.

\subsubsection{\textbf{Distributed Gradient Computation}}
Leveraging the serverless concurrent abilities of AWS Lambda, we implement a parallel gradient computation within our architecture. Each peer partakes in this distributed process by calculating gradients on assigned data batches and storing these computed gradients in its local Redis database. To expedite the gradient computation, data, segmented into smaller batches, is fetched from S3 storage, and model parameters are retrieved from Redis.

\subsubsection{\textbf{Averaged Gradient Computation}}

Once the gradients have been computed, an embedded Lua script calculates the gradients' average within the Redis database environment, capitalizing on in-database programming. This approach eliminates costly external data transfers. Once local averaging is complete, peers send a completion message to the ''sync queue``, notifying others of the task's conclusion.

\subsubsection{\textbf{Heartbeat Monitoring}}

Our system incorporates a 'heartbeat' mechanism designed to be triggered every epoch, where each peer checks the operational status of other peers databases. 

Peers send a signal, waiting for responses to confirm others' activeness. Failure to respond within a set timeframe and a number of trials denotes a peer as ''inactive``, subsequently removed from the trusted peers list and added to the ''inactive peers info`` list. This continuous health check is vital for network integrity and uninterrupted peer-to-peer communication.

\subsubsection{\textbf{Peers Synchronization}}

To collate the individually computed average gradients from each participating peer and derive the aggregated gradient, a specially designated Lambda function ''synchronize`` is instantiated. This function serves as a synchronization barrier that waits until the count of messages in the queue equals the current number of active peers, those determined by the preceding heartbeat check. 

\subsubsection{\textbf{Gradient Aggregation}}

Once all peers are synchronized, the gradient aggregation phase begins. Each peer fetches the average gradients from the databases of all active peers in the network. An aggregation function then amasses these gradients, utilizing robust algorithms to discard outlier gradients. The final aggregated gradient is then stored in each peer's Redis database. This approach ensures the integrity and accuracy of our gradient aggregation process.

\subsubsection{\textbf{Model Update}}

Utilizing the in-database programming feature of RedisAI, we directly update each peer's model parameters stored in the Redis database using the aggregate gradient, thereby bypassing the conventional read-process-write cycle.

\subsubsection{\textbf{Convergence Checking}}

Upon the completion of model parameter updates, a Lambda function is intermittently called (e.g., after every tenth iteration) to check model convergence. This approach reduces unnecessary function calls, as significant model changes aren't anticipated after each iteration.

\subsubsection{\textbf{Update and Trigger new epoch}}

In the context of our AWS Step Function based workflow, each step function orchestrates the process for a single epoch. The stateful nature of this service necessitates the instantiation of a new Step Function at the end of each epoch to maintain the continuity of training inside each peer. To facilitate this, we employ a dedicated Lambda function that is responsible for spawning the new Step Function. This Lambda function initializes the new Step Function with the correct inputs, including the subsequent epoch number, the degree of parallelism for gradient computation tasks, and a flag indicating whether a convergence check is required (based on the current epoch number). This process ensures an updated step function based on current needs. The ARN (Amazon Resource Name) of the new function is stored in the Redis database, which is reused in subsequent epochs if no new requirements arise.

The same Lambda function also undertakes the responsibility of updating the list of inactive peers through a consensus-driven approach. To ensure the integrity of this process, the list of inactive peers for each node is cross-validated against corresponding lists from all other nodes within the network. An inclusive agreement is applied, meaning a peer is only marked as inactive if it is listed as such in every peer's record.

\subsubsection{\textbf{Fault Tolerance and Peer Data Redistribution}}

A core strength of our architecture lies in its fault-tolerance mechanisms. This resilience is primarily manifested through our approach to handle instances of peer inactivity or failure. When one or many peers become inactive, our dedicated Lambda function first identifies these instances using the consensus-based approach described previously. Once inactive peers are identified, the data originally assigned to the downed peers is segmented and distributed among the active peers based on a predefined ranking system. Here, each peer, according to their rank, inherits a corresponding portion of the data from the inactive peer.

Following this data reassignment, the ''Update and Trigger new epoch`` Lambda function adjusts the configuration of the new Step Function to account for the change in data distribution and the increased workload for each active peer. This adjustment may include tuning the degree of parallelism for gradient computation tasks to accommodate the additional data batches.

\section{Experimental Setup}
\label{sec:setup}

In order to evaluate the efficacy of the Serverless Peer Integrated for Robust Training (SPIRT) architecture, we design a series of experiments to evaluate the performance of various CNN models across different datasets on the proposed architectures.

\subsection{Datasets}

We utilized two public datasets for our experiments:

\textbf{MNIST:} The MNIST Handwritten Digit Collection \cite{deng2012mnist} consists of 60,000 handwritten digit samples, each belonging to one of ten classes.

\subsection{Model Architectures and Hyperparameters}

The experiments involve three different CNN models:

\textbf{MobileNet V3 Small:} A lightweight CNN developed for mobile and edge devices, it features inverted residual blocks, linear bottlenecks, and squeeze-and-excitation modules, with roughly 2.5 million trainable parameters \cite{koonce2021mobilenetv3}.

\textbf{ResNet-18:} A deep learning CNN model with approximately 11.7 million parameters, featuring 18 layers and using "skip connections" to aid training of deeper networks \cite{he2016deep}.

\textbf{DenseNet-121:} A uniquely structured CNN with about 8 million parameters and 121 layers. It leverages dense connections, where each layer is connected to all other layers, enhancing learning efficiency \cite{huang2017densely}.

\subsection{Redis and RedisAI Configuration}

We utilized Redis and RedisAI for in-database model updates. Each peer's Redis was deployed within an Amazon EC2 R6a.large instance. To access the database hosted on EC2, we utilized SSH tunnels to forward the port and enable access from public services such as AWS Lambda.

\subsection{AWS Lambda Configuration}
Several AWS Lambda functions, discussed in Section \ref{sec:study-design}, were set up for the training procedure. Dependencies such as PyTorch, NumPy, Redis, RedisAI, and sshtunnel were necessary. Managing these dependencies posed a challenge due to AWS Lambda's deployment package size limit. The zipped package must be less than 50MB, and the unzipped files cannot exceed 250MB.

We utilized a Virtual Private Cloud (VPC) to minimize the need for multiple SSH connections between compute gradients Lambda functions and the database EC2 instance.



\begin{figure*}[ht]
    \centering
    \begin{minipage}{\textwidth}
        \centering
        \subfloat[MobileNet V3 Small Model\label{fig:varying_batch_mobilenet}]{{\adjustbox{height=4cm,width=8.5cm}{\includegraphics[width={0.35\linewidth}]{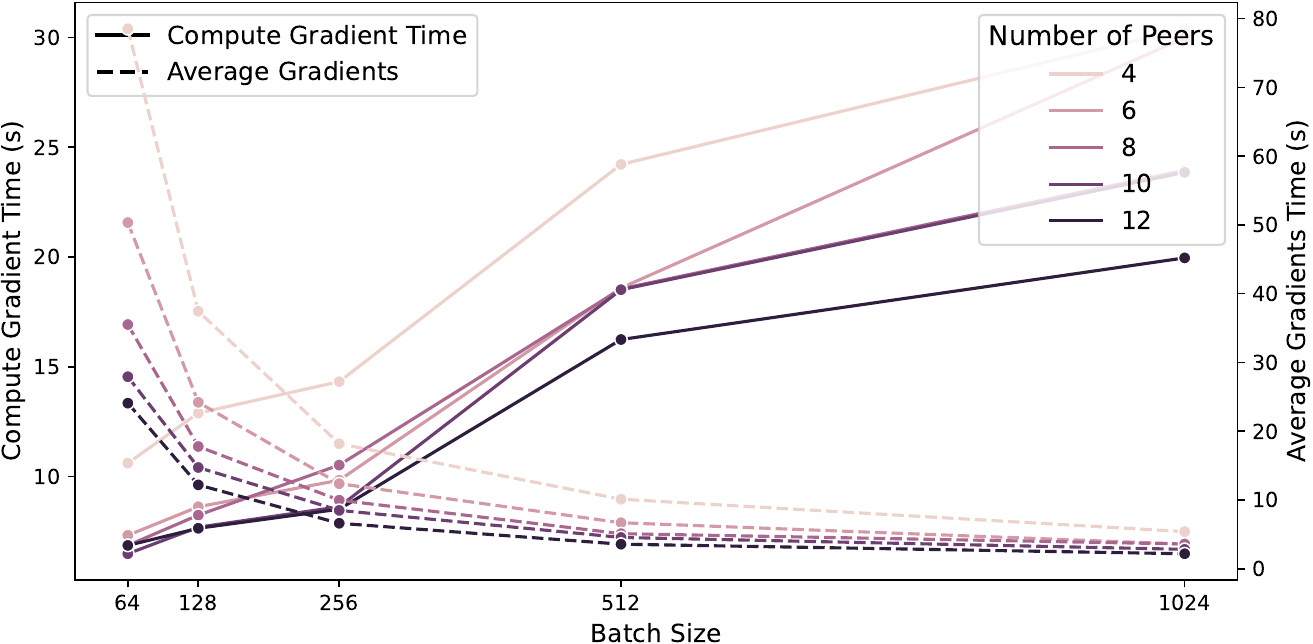}} }}%
            \hspace{-0.3cm}
        \quad
        \subfloat[DenseNet-121 Model\label{fig:varying_batch_dense}]{{\adjustbox{height=4cm,width=8.5cm}{\includegraphics[width={0.35\linewidth}]{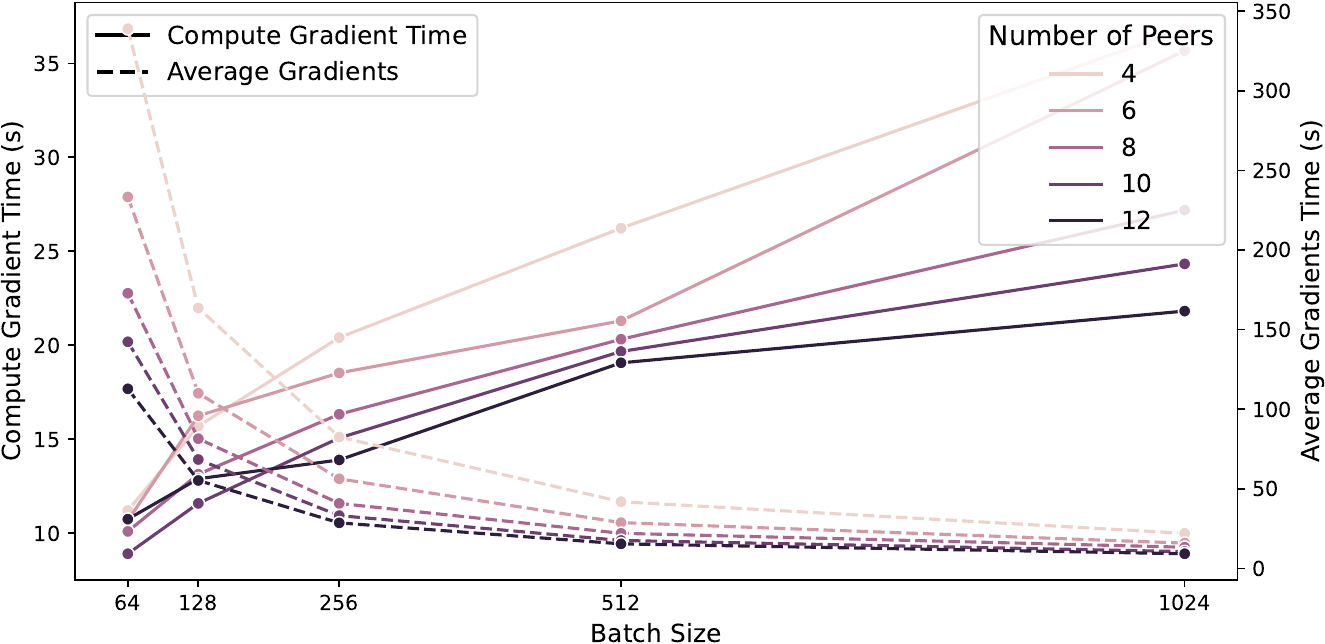}} }}%
        \caption{Comparison of Compute Gradient Time and Average Gradients Across Varying Batch Sizes for Different Number of Peers}%
        \label{fig:varying_batch}%
    \end{minipage}%
    \vspace*{-10pt}
\end{figure*}

\section{Efficiency and Performance in Serverless P2P Architecture}
\subsection{Motivation:}
Our innovative serverless peer-to-peer architecture marks the advent of a decentralized era in model training. Peers are enabled to concurrently compute gradients in parallel based on their assigned datasets, exchange these gradients for robust aggregation, and update their model parameters. Yet, a theoretically sound architecture is merely the initial phase; the critical next step is to assess its performance under various operational conditions using multiple deep learning models of different complexity and computational demands.

Moreover, it's crucial to explore scalability in two significant aspects: \textit{(1) Intra-peer scalability:} The ability of our architecture to manage parallel gradient computations within a single peer, and \textit{(2) Inter-peer scalability:} The adaptability of our architecture to changing numbers of peers in the system.

Our motivation lies in thoroughly understanding our architecture's scalability under these two critical dimensions. By altering the number of peers and harnessing each peer's parallelization capabilities, we can observe their impacts on performance metrics like gradient computation time, aggregation time, and overall training time per epoch. This approach gives us a holistic view of our architecture's performance under varied operational circumstances.

\begin{figure}[h!]
 \centering
 \includegraphics[scale=0.45]{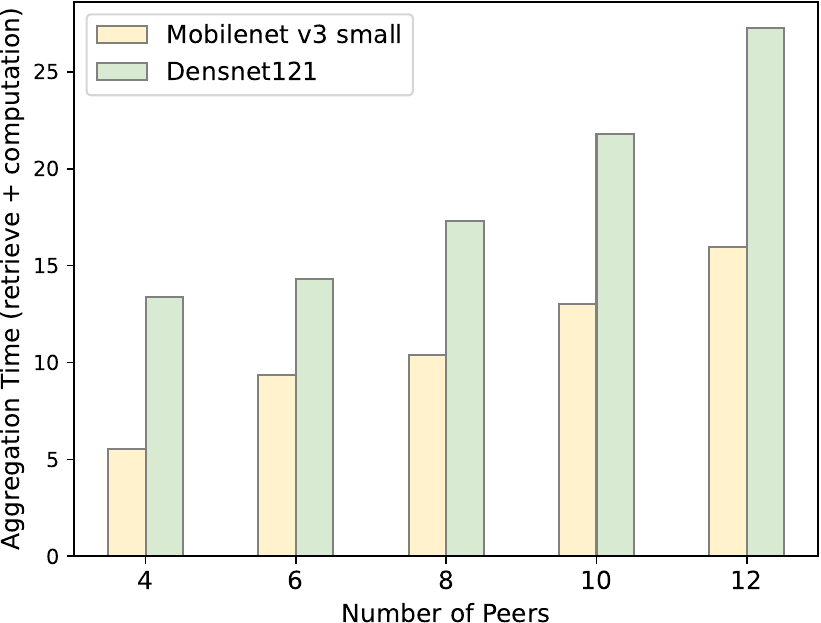}
 \caption{Comparison of Average Aggregation Times for Mobilenet v3 small and Densnet121 Models Against \# of Peer}
 \label{fig:aggregation}
 \vspace*{-10pt}
\end{figure}

\subsection{Approach:} For our experiments, we utilized two distinct deep learning models, Densenet121 and MobileNet V3 Small, to represent a range of model complexities, using the standard MNIST dataset for training.

\textit{Intra-peer scalability} was explored by adjusting the batch size, thereby determining the number of concurrent gradient computations within a peer. With batch sizes of 64, 128, 256, 512, and 1024, we traced the impact on gradient computation and averaging times.

\textit{Inter-peer scalability} was assessed by varying the number of peers, starting from 4 and incrementing by 2, up to 12. This helped us gauge the system's adaptability and its effect on parameters like aggregation time and training time per epoch.

\subsection{results:}

\begin{figure*}[h!]
    \centering
    \begin{minipage}{.7\textwidth}
        \centering
        \subfloat[MobileNet V3 Small Model\label{fig:parameter-serverq}]{{\adjustbox{height=4cm,width=6cm}{\includegraphics[width={0.35\linewidth}]{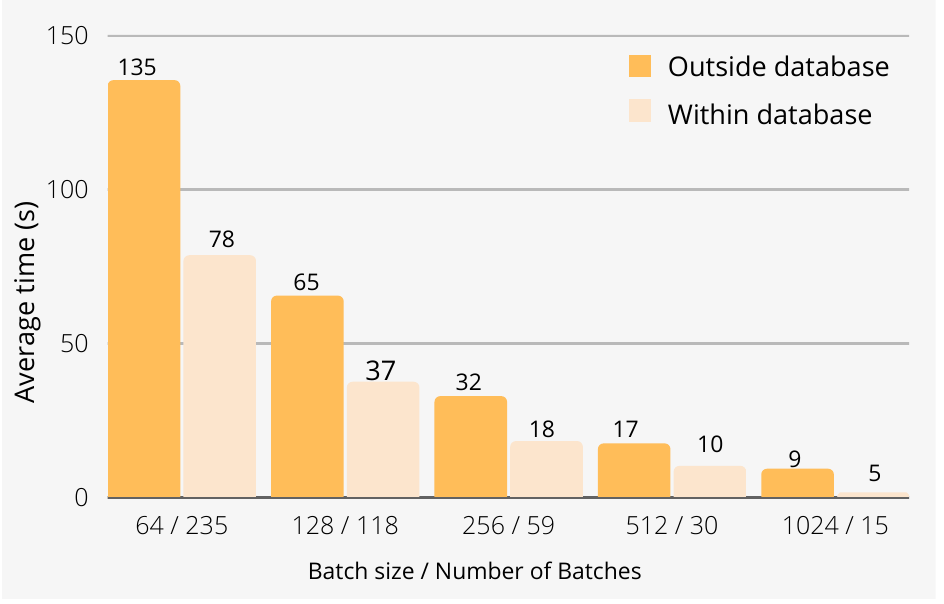}} }}%
            \hspace{-0.3cm}
        \quad
        \subfloat[Resnet-18 Model\label{fig:p2p1}]{{\adjustbox{height=4cm,width=6cm}{\includegraphics[width={0.35\linewidth}]{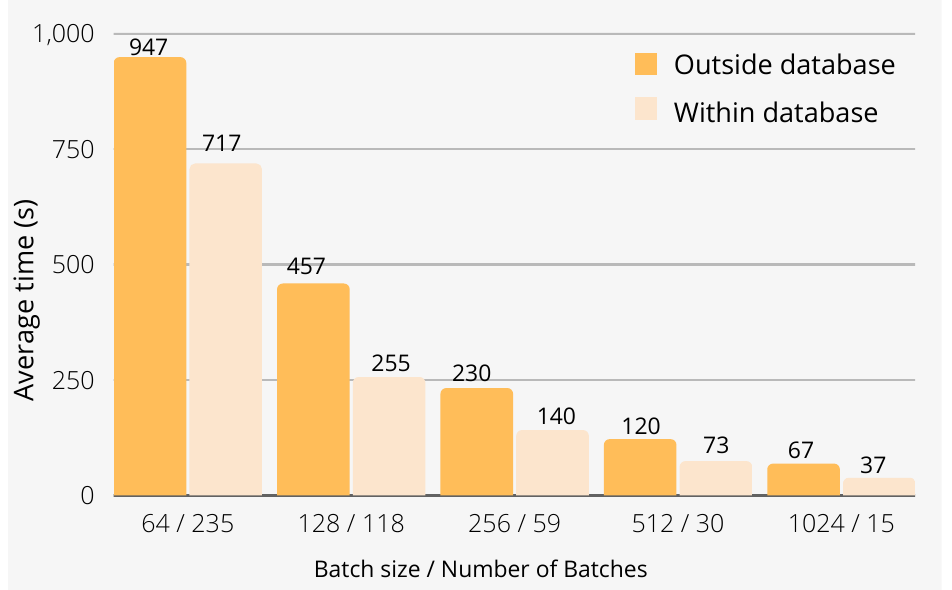}} }}%
        \caption{Time taken for calculating gradient averages within and outside the database}%
        \label{fig:time_peers1}%
    \end{minipage}%
    \begin{minipage}{.3\textwidth}
        \centering
        \adjustbox{height=4cm}{\includegraphics[width={0.45\linewidth}]{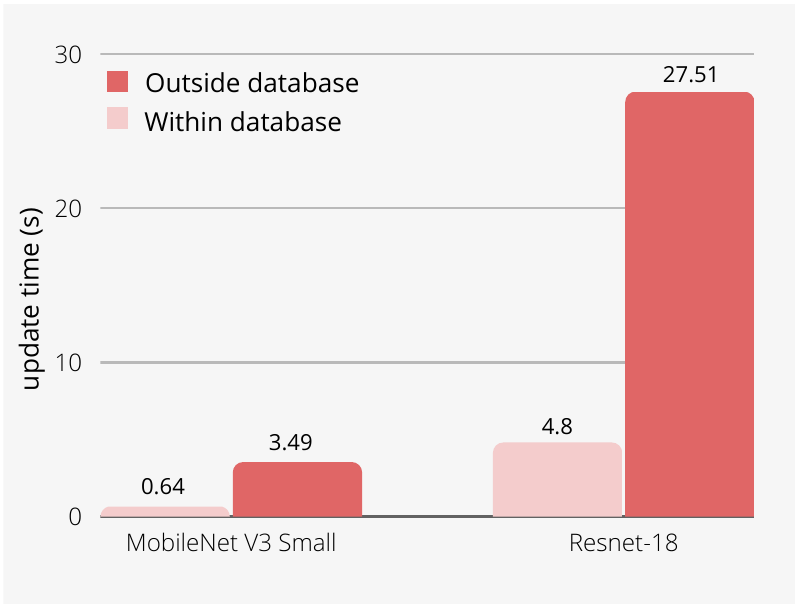}}

        \caption{Time taken for Model Update within and outside the database}%
        \label{fig:time_peers2}%
        \hspace{-0.3cm}
    \end{minipage}
    
    \vspace*{-10pt}
\end{figure*}

As we observed in Figure \ref{fig:varying_batch}, an increase in batch size yields an increased time to compute gradients. This is expected, as a larger batch size implies more data to process at once per peer, which intuitively would increase the time needed for computation. We observed this increase in computation time consistently across both Densenet121 and MobileNet V3 Small models, signifying that the effect is model-agnostic. Additionally, this trend persists across different numbers of peers. This indicates that the increase in computational burden due to a larger batch size cannot be offset simply by adding more peers, as the computation time per gradient computation still increases.
\begin{table}[h]
\centering
\caption{Training Time per Epoch Across Different Batch Sizes and Peer Counts for MobileNet V3 Small and Densenet121 Models}

\begin{adjustbox}{max width=8cm}
\label{tab:combined}
\begin{tabular}{|c||c|c|c|c|c||c|c|c|c|c|}
\hline
& \multicolumn{5}{c||}{Mobilenet v3 small (Batch Size)} & \multicolumn{5}{c|}{Densenet121 (Batch Size)} \\ 
\hline
\# Peers & 64 & 128 & 256 & 512 & 1024 & 64 & 128 & 256 & 512 & 1024 \\ 
\hline
4  & 96.27 & 57.46 & 39.61 & 41.41 & 42.85 & 375.82 & 202.62 & 125.7  & 90.67 & 81.17 \\
6  & 68.55 & 43.69 & 33.05 & 36.11 & 44.39 & 268.13 & 150.16 & 98.41  & 73.39 & 73.77 \\
8  & 54.42 & 38.05 & 32.5  & 35.64 & 39.48 & 209.79 & 121.41 & 83.65  & 67.44 & 65.25 \\
10 & 48.97 & 36.97 & 31.65 & 37.56 & 41.17 & 182.32 & 110.95 & 79     & 67.02 & 63.89 \\
12 & 48.43 & 37.3  & 32.57 & 37.21 & 39.53 & 169.12 & 104.39 & 74.41  & 65.19 & 62.03 \\
\hline
\end{tabular}
\end{adjustbox}
\end{table}

On the other hand, as we decrease the batch size, we enable more parallel gradient computations due to the availability of more smaller-sized batches. Although this parallelization could potentially lead to faster computation times, we interestingly notice an increase in the average computation time within each peer's database due to the overhead associated with averaging multiple gradients.

Further dissection of the number of peers' impact on performance, independent of batch size, reveals an interesting trend. As illustrated in Figure \ref{fig:aggregation}, the aggregation time, which includes both the computation of the aggregated gradient and the time to retrieve locally averaged gradients from other peers, decreases when the number of peers is reduced (as shown in the bar plot). This suggests an inherent computational overhead linked with gradient aggregation over larger networks of peers, a factor that requires careful consideration when designing and scaling such decentralized learning systems.


This network size impact is further reflected in Table \ref{tab:combined}, which present the total training time per epoch for both Densenet121 and MobileNet V3 Small models, respectively. We observe an overall decrease in training time per epoch with a larger batch size and a higher number of peers. Nonetheless, these tables also highlight the nonlinear relationship and the trade-offs between batch size and peers network scale.
\section{Optimizing Communication Overhead}
\subsection{Motivation:} 
As we delve deeper into the intricate world of distributed serverless environments, characterized by numerous autonomous and stateless services, the challenges associated with frequent database retrieval loads during training phases emerge. While works such as previous work \cite{ic2e2023_barrak}, LambdaML \cite{P46}, SMLT \cite{P54}, and MLLess \cite{sarroca2022mlless} utilize a database as a communication channel, storing and retrieving model parameters as necessary, the exploration of the communication overhead these operations create remains largely unexplored. Such conditions often precipitate a significant increase in communication overhead, consequently undermining the overall training performance. However, within our unique architectural framework, we incorporate a customized Redis. This component allows us to perform average and update operations directly within Redis. Through this investigation, we aim to shed light on how our approach can reduce communication overhead and subsequently enhance training performance.

\subsection{Approach:}
In our experiment, we first explore the efficiency gains that can be realized by conducting model updates directly within RedisAI. Following this, we delve into the benefits of calculating gradient averages within Redis. For a comprehensive evaluation that considers the impact of model size on the overhead, we will use two distinct models - MobileNetV3 Small and ResNet18 - and run these models on the MNIST dataset. The insights from these tests will then be compared with the traditional method of iterative fetch-update-store operations, typically used with standard Redis, as outlined in previous studies.

\subsection{Results:}
\begin{figure*}[h]
    \centering
    \subfloat[No attack\label{fig:healthy}]{{\includegraphics[width={0.3\textwidth}]{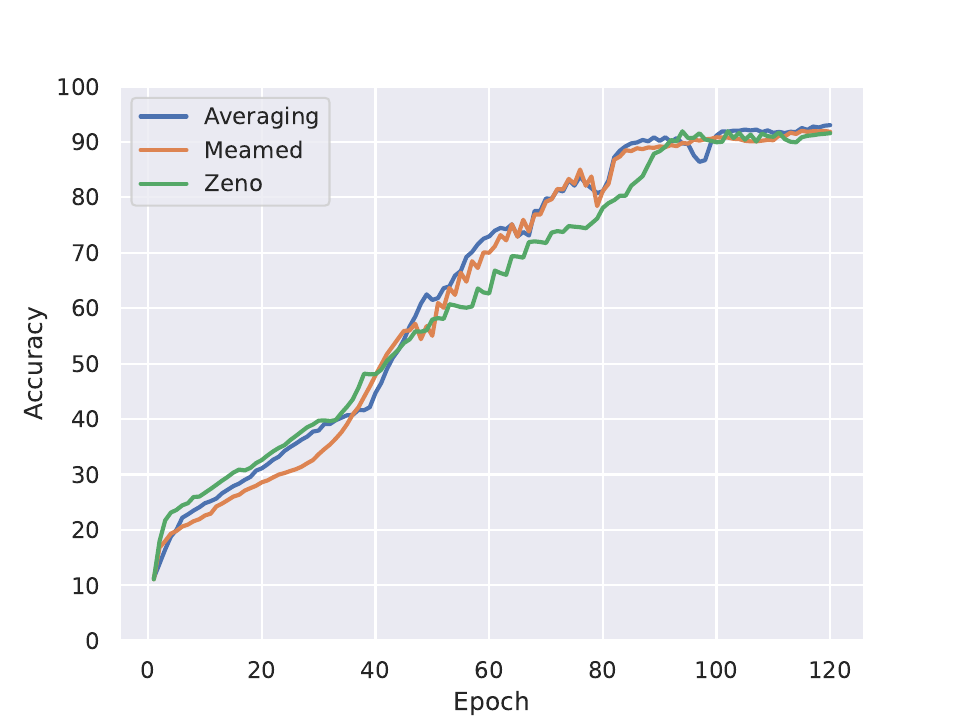} }}%
    \qquad
    \subfloat[Sign Flip Attack\label{fig:sign}]{{\includegraphics[width={0.3\textwidth}]{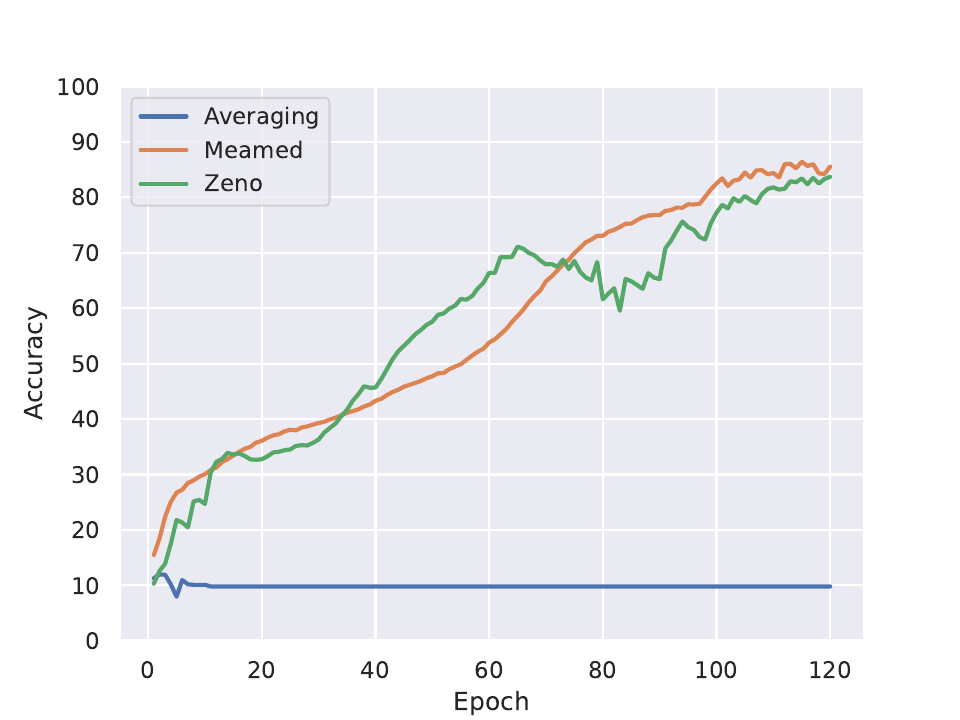} }}%
    \qquad
    \subfloat[Noise Attack\label{fig:noise}]{{\includegraphics[width={0.3\textwidth}]{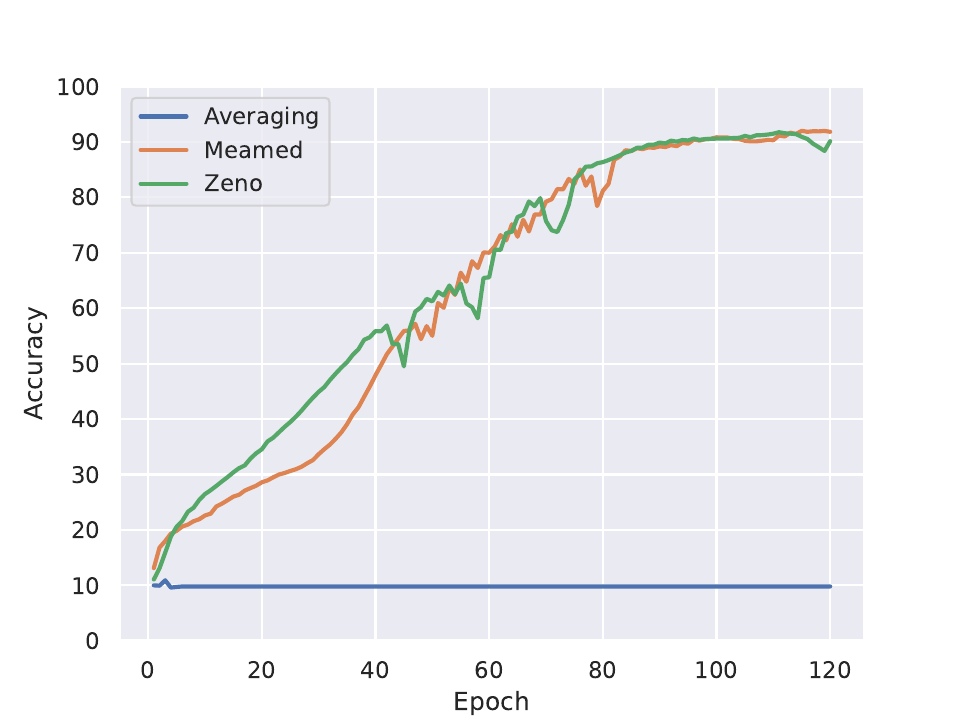} }}%
    \caption{Evaluation of network accuracy using Averaging, Zeno, and Meamed aggregation methods under various conditions: (a) Normal training with no attack, (b) Training under sign flip attack, where gradients are manipulated, and (c) Training amidst a noise attack, where random Gaussian noise is added to local updates.}%
    \label{fig:time_peers}%
    \vspace*{-10pt}
\end{figure*}

An in-depth analysis of the experimental results provides persuasive evidence of the significant efficiency improvements facilitated by in-database operations, as vividly depicted in Figures \ref{fig:time_peers1} and \ref{fig:time_peers2}.

Figure \ref{fig:time_peers1} concisely illustrates the substantial reduction in time for gradient averaging calculations within the database, as opposed to outside, for both the MobileNetV3 Small and the ResNet-18 models. As the plot underscores, the in-database approach yields consistently lower computation times across the spectrum of batch sizes. The MobileNetV3 Small model, for example, experiences an impressive decrease in computation time from 135.29 seconds outside the database to 78.52 seconds within, for a batch size of 64. The time efficiency continues to escalate as the batch size increases, with the largest batch size of 1024 resulting in a striking 82\% improvement, requiring only 5.4 seconds for in-database computations.

A similar efficiency advantage is observed with the larger ResNet-18 model, where gradient averaging computations conducted within the database halved the processing time to 37.41 seconds, a stark reduction from the 67.32 seconds required when computed outside the database, for the batch size of 1024. This trend equates to a remarkable 69\% improvement in processing efficiency, proving that our approach's benefits are applicable even for larger, more demanding models.

Figure \ref{fig:time_peers2} delves into the time taken for model updates within and outside the database. It echoes the efficiency improvements demonstrated by the previous figure, particularly the sharp decrease in model update times when processed within the database. MobileNetV3 Small model updates saw an 82\% reduction in time when executed within the database, shrinking from 3.49 seconds outside the database to a mere 0.64 seconds within. Likewise, ResNet-18 updates experienced an approximately 83\% improvement, with update times reducing dramatically from 27.5 seconds outside the database to 4.8 seconds within.

\section{Scalability and Fault Tolerance Evaluation}

\subsection{Motivation:} In the dynamic realm of distributed machine learning (ML) peer-to-peer training, disruptions, such as peer failures, new peers joining, and potential Byzantine behaviors, are inevitable, and if not effectively managed, can impact the reliability and performance of the ML training. This provides the impetus for our experiment, which is meticulously designed to rigorously evaluate the robustness, scalability, and fault tolerance of a serverless, peer-to-peer ML training architecture. In this architecture, peers are carefully identified to form a secure training network, and its capacity to handle peer failures, seamlessly integrate new peers, and resist Byzantine attacks will be critically evaluated to ensure long-running distributed ML training sessions can continue to deliver accurate results amidst these disruptions.
\subsection{Approach:}

In our experimental approach, we center on the evaluation of three crucial aspects using the Mobilenet V3 Small model trained on the MNIST dataset: \textit{peer failures}, the \textit{addition of new peers}, and \textit{Byzantine attacks}.

We initiate with the simulation of \textit{\textbf{peer failures}} where we begin training with four peers, each processing 15 compute gradients. A peer failure is then artificially induced, leading us to measure the time taken by the remaining peers to identify the failure. In response to this failure, these remaining peers incorporate the data of the failed peer into their following epochs, augmenting the compute gradients to 20. This simulation aids in testing the system's resilience and ability to adapt in response to computational loss during peer failures.

To assess the system's scalability, we execute the \textit{\textbf{addition of new peers}} scenario where a new peer is introduced into the network. We proceed to calculate the time taken by the existing peers to recognize and integrate this newcomer into the ongoing training process.

astly, our approach includes countering potential \textit{\textbf{Byzantine attacks}} using robust aggregation algorithms, Zeno and Meamed. 
Zeno \cite{xie2019zeno} ensures resilience by using a validation set to score and exclude any potentially adversarial local updates. The medians-based approach, or Meamed  \cite{xie2018generalized}, combats Byzantine attacks by creating a vector that minimizes the overall distance to all local updates, thereby mitigating the influence of adversarial elements. By simulating adversarial scenarios of one malicious peer, such as a sign flipping attack \cite{li2019rsa} where the malicious peer inverts and amplifies its local gradient, and a noise attack \cite{li2020learning} where the malicious peer introduces Gaussian noise to its local updates, we track the training progression to convergence. This comprehensive approach enables us to gauge the architecture's resilience and robustness against adversarial behavior.

\subsection{results:}

\subsubsection{\textit{\textbf{Peer Failure}}} 
\begin{figure}[h!]
 \centering
 \includegraphics[scale=0.3]{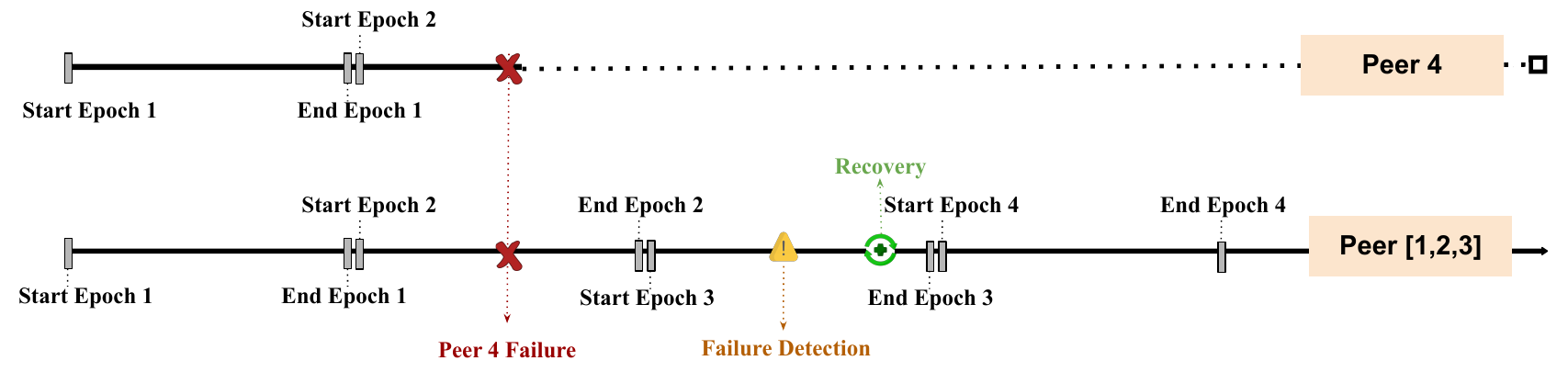}
 \caption{Recovery Time when a Peer Fails}
 \label{fig:peer_failure}
\end{figure}

The flow of the experiment are depicted in Figure \ref{fig:peer_failure} for a more visual understanding.
Initially, we began with four peers, each designed to process 15 batches per epoch. The first epoch proceeded unimpeded, with the total training time recorded at 52.6 seconds.

A simulated peer failure has been introduced at the beginning of the second epoch, immediately after a health check had validated the 'failed' peer as operational. This timing represented a worst-case scenario, as it extended the detection period. Nonetheless, the remaining peers were in recognizing the failed peer within the ongoing epoch, taking a total 50.9 seconds to align with the next epoch's heartbeat step.

Post this single-peer-level detection, the remaining peers reached a consensus on the failed peer within 9.66 seconds. This aggregated the total detection time to 61.56s seconds of the failed peer.  
we note that, despite the deferred detection of the simulated failure in the third epoch, the training process was not interrupted, thus maintaining the progression towards model convergence. 
Upon consensus on the peer failure, the recovery process was triggered. This involved creating a new AWS Step Function to redistribute the failed peer's computational workload, which increased the remaining peers' load from 15 to 20 batches per epoch.
The complete recovery process, encapsulating the creation, deployment, and database entry of the new AWS Step Function, required an extra 1.94 seconds. After recovery, the remaining peers continued training in the 4th epoch with an adjusted workload to account for the lost peer.
By the end of the fourth epoch, the total training time post-recovery was registered as 55,06 seconds.

Our analysis shows a slight increase in total training time from 52.6 to 55.06 seconds, as the system transitions from four to three peers. This can be attributed to the increased number of gradients being averaged due to the redistribution of the failed peer's workload.

\subsubsection{\textit{\textbf{Adding a new peer}}}
We found that in the scenario of \textit{adding a new peer}, it took approximately 7.2 seconds for the existing three peers to recognize and integrate the new peer into their list of trusted peers.

\subsubsection{\textit{\textbf{Tolerating Byzantine attacks}}}

The adoption of Meamed and Zeno  aggregation algorithms introduced a computational overhead, leading to an approximate increase of 8.2 and 5.9 times in computational time, respectively, in contrast to the average aggregation method.

In a scenario without adversarial attacks, all three aggregation methods – Averaging, Zeno, and Meamed – achieved an accuracy above 90\% within approximately 100 epochs. During a sign flip attack, the robust aggregations, Zeno and Meamed, managed to converge to almost 85\%, while the normal averaging method did not. During a noise attack scenario, both Zeno and Meamed reached convergence above 90\% after nearly 90 epochs, but the normal averaging method remained divergent.

\section{Discussions}
\label{sec:discussions}
This section critically examines the results of our study and the performance of SPIRT, identifying its implications for peer-to-peer serverless ML training and potential directions for future research.

\subsection{Serverless P2P for ML training}
In our research, we created a distributed training workflow utilizing serverless computing, customized to inherit characteristics from a peer-to-peer (P2P) system, such as robustness, fault tolerance, and scalability. This design led to the creation of logical peers, each focusing on training a distinct part of the dataset. The statelessness of serverless computing necessitates the use of redis databases into our architecture for maintaining state information and enabling inter-component communication. To reduce communication overhead, we incorporated a modified version of RedisAI into our architecture, enabling immediate model updates within the database. Our proposed architecture paves the way for further exploration into the enhancement of in-database computing in distributed serverless ML systems.

\subsection{Security, reliability, and Fault Tolerance in Machine Learning Architecture}

Exploring scalability, a cornerstone of distributed systems, we delved into balancing inter and intra-peer scalability. While handling multiple data batches within a single peer increases efficiency, it also risks a single point of failure. Conversely, extending processing power across multiple peers introduces challenges related to gradient averaging and multiple connections. By testing our system with varied model and batch sizes, and different numbers of peers, while further exploration is required to fully comprehend these trade-offs, our efforts lay a significant foundation towards improving scalability in distributed ML systems.

In terms of fault tolerance, our design incorporates a robust heartbeat system to monitor peer health. Although this method is generally effective, it can introduce minor delays in detecting failures due to its periodic checking nature. To enhance detection speed, more frequent heartbeats could be implemented. As for fault recovery, our architecture ensures swift recovery time either  by intra-peer scalability (adding a new peer) or by redistributing the load among existing peers (inter-peer scalability).

Aiming to augment the security of our architecture, we've implemented sophisticated cryptographic mechanisms which provide robust safeguards for data integrity, authenticity, and confidentiality. These measures ensure that data shared between two peers can only be understood by them. In addition, we've adopted robust aggregation that securely consolidates gradients from diverse peers. While the robust aggregation procedure may extend the aggregation time - for instance, take longer than the average aggregation - its capacity to safeguard against Byzantine attacks and guarantee model convergence underscores its importance.



\section{Conclusion}
\label{sec:conclusion}




In this study, we present SPIRT, a serverless machine learning (ML) architecture that streamlines training in distributed settings. By minimizing communication overhead, demonstrating resilience to adversarial attacks, and seamlessly integrating new peers.

We found that SPIRT's use of in-database operations in RedisAI resulted in remarkable computational and communication efficiency, improving processing times by up to 82\%. Moreover, SPIRT exhibited a high degree of resilience amidst disruptions, rapidly adapting to peer failures with minimal impact on training times and integrating new peers swiftly into the training network. When subjected to adversarial scenarios, the robustness of the architecture became more apparent. Through the use of Byzantine-resistant aggregation methods, Zeno and Meamed, high accuracy levels above 85\% and 90\% were maintained during sign flip and noise attacks, respectively.

These findings provide a robust foundation for the evolution of distributed ML training frameworks, paving the way for advancements in efficiency, resilience, and scalability in serverless environments.

\balance
\bibliographystyle{IEEEtran}
\bibliography{references} 
\end{document}